\documentclass[11pt,twoside]{article}
\usepackage{asp2014}

\aspSuppressVolSlug
\resetcounters

\begin{document}

\title{Application of Google Cloud Platform in Astrophysics}

\author{Marco~Landoni,$^1$, G. Taffoni$^2$, A. Bignamini$^2$ and  R. Smareglia$^2$}
\affil{$^1$Istituto Nazionale di Astrofisica, Osservatorio Astronomico di Brera, Via E. Bianchi 46, Merate (LC) - ITALY; \email{marco.landoni@inaf.it}}
\affil{$^2$Istituto Nazionale di Astrofisica, Osservatorio Astronomico di Trieste. Via G. B. Tiepolo, Trieste - ITALY}           % remove/add authors as you need

\paperauthor{Marco~Landoni}{marco.landoni@inaf.it}{ORCID}{ISTITUTO NAZIONALE DI ASTROFISICA}{Author1 Department}{City}{State/Province}{Postal Code}{Country}
% remove/add authors as you need
\paperauthor{Sample~Author2}{Author2Email@email.edu}{ORCID_Or_Blank}{Author2 Institution}{Author2 Department}{City}{State/Province}{Postal Code}{Country}

% leave these next few aindex lines commented for the editors to enable them
%\aindex{Landoni,~M.}
%\aindex{Coauthor,~A.}          % remove and add as you need
  
\begin{abstract}

The availability of new Cloud Platform offered by Google motivated us to propose nine Proof of Concepts (PoC) aiming to demonstrated and test the capabilities of the platform in the context of scientifically-driven tasks and requirements.
We review the status of our initiative by illustrating 3 out of 9 successfully closed PoC that we implemented on Google Cloud Platform. In particular, we illustrate a cloud architecture for deployment of scientific software as microservice coupling Google Compute Engine with Docker and Pub/Sub to dispatch heavily parallel simulations. We detail also an experiment for HPC based simulation and workflow executions of data reduction pipelines (for the TNG-GIANO-B spectrograph) deployed on GCP. We compare and contrast our experience with on-site facilities comparing advantages and disadvantages both in terms of total cost of ownership and reached performances.
  
\end{abstract}

\section{Introduction}

Google Cloud Platform (GCP) offers a variety of services, ranging from storage to high performance computing and workflow execution, that could be exploited in the context of Computational Astrophysics. In this paper we review three Proof of Concept (PoC) out of the nine proposed to Google that have been successfully implemented on the public platform illustrating the architecture and the main results we have obtained. The paper is organized as follows: in Section 2 we illustrate the PoC for HTC oriented application while in Section 3 we comment on the implementation of HPC cluster on the Google Cloud Platform. We also tested the execution of Workflows  aiming to offer instrument pipeline as a service reporting the results in Section 4.

\section{POC 1 - HTC Workload on Google Cloud Platform. The case of DIAMONDS}
\begin{figure}
  \includegraphics[scale=0.22]{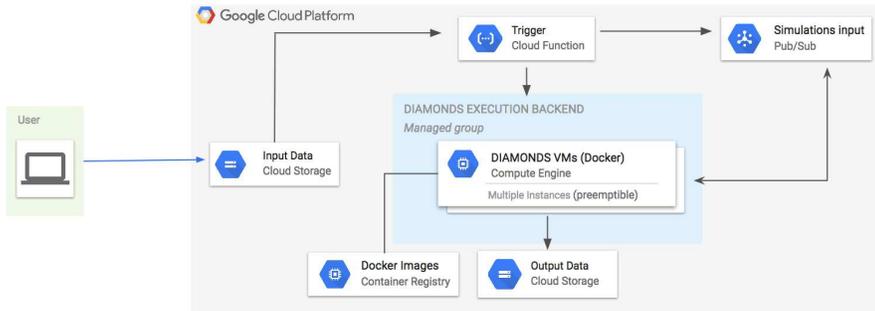}
  \caption{The HTC serverless architecture for DIAMONDS pipeline (POC1)}
  \label{fig:diamonds}
\end{figure}

DIAMONDS \citep{corsaro} is a Bayesian inference code that is design to process data from asteroseismology, a technique that study stars oscillations through photometry or spectroscopy in order to derive their internal structure and physical parameters, such as the true mass. DIAMONDS has demonstrated (on premises) to be runnable in parallel through \textit{embarrassingly parallelism} paradigm with almost no network communication. This kind of computational approach is very suitable to test HTC workloads on GCP. We introduce in this scenario a concept of \textit{serverless HTC scheduler} that fruitful exploit an heterogeneous set of GCP components such as Pub/Sub, Cloud Functions and Managed Groups in GCE (see Figure \ref{fig:diamonds} and \cite{landoni} for further implementation and details). Typical resource schedulers are complex middleware that have to deal with a limited amount of resources available accordingly to a set of time-sharing policies. An advantage of this approach is to use some equivalent concepts such as the queue from the available services to manage the execution of HTC workloads while guaranteeing a general purpose approach to many possible HTC computation. In the architecture that we design for GCP, keeping in mind to be as much as general as possible, the computation starts by uploading to Cloud Storage a plain text file that contains, for each row, the data necessary to perform a single run. These rows are pushed, by a triggered Cloud Function, into a Pub/Sub topic. Then, a cluster of instances (Regular or Preemptible and configured using Google Managed Groups) dimensioned runtime is fired up accordingly to the estimated size of the whole workload. Each node of the cluster, after starting up with a pre-configured Image on Compute Engine, pulls a number of messages (proportional to the number of vCPUs available) from the PubSub queue starting the computation of various DIAMONDS simulations using Docker containers. Data produced locally by DIAMONDS on each instance are finally transferred to a bucket on Google Cloud Storage before shutting down.  This method allows to deploy an HTC-based architecture, suitable for many projects that share the same kind of parallelism and requirements on the workload, that scales both vertically (number of cores per node and thus number of simulations) and horizontally through an elastic cluster fired up accordingly to the number of required simulations and CPU/hours.

\section{POC 2 - Exploring HPC capabilities with Google Cloud Platform}
GADGET  \citep{gadget} is a lagrangian code to perform numerical simulations of gravitationally interacting particles of both dark matter and baryonic matter which  computes gravitational forces using a TreePM technique. A mean field approximation is used for large scales (Particle-Mesh, PM) while at smaller scales a usual Treecode is used. Hydrodynamics is solved using a so-called Smoothed Particle Hydrodynamics technique. GADGET is an HPC code based on message passing interface (MPI) libraries and OpenMP. It is written in C and requires some support libraries to run (FFW2.4, HDF5, GSL). To test the performance of the Google Cloud infrastructure for HPC
applications, we executed a virtual cluster managed by Slurm scheduler composed by both reserved nodes and On-demand (non-preemptible) instances. To automate the deployment of the cluster, we use the Cloud Deployment Manager (CDM). This service aims at automate the creation of complex resources and services where various entities are described in terms of \textit{yaml} files and deployed using \textit{gcloud} command line interface. In this POC, we modified the Slumr official CDM files to modify the resources (we used 4 cores with 4 GB ram per core) and the software (we added MPI, FFTW, HDF5, and GSL). Moreover, we deployed a cluster where only two computing nodes are running and more
resources are bootstrapped on demand using Slurm only if necessary. 
\begin{figure}
  \includegraphics[scale=0.35]{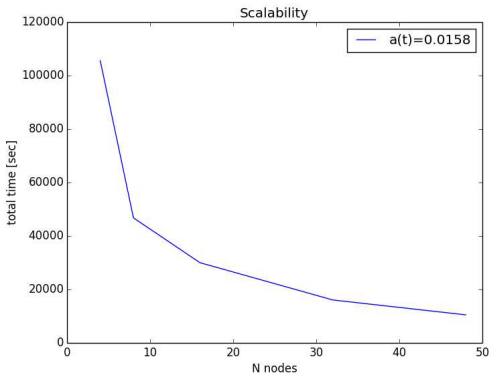}
  \caption{GADGET sclalability on GCloud.}
  \label{fig:gad2}
\end{figure}
We tested GADGET scalability with a small cosmological BOX of 778688 particles for a $\Lambda$CDM model ($\Omega_0$ = 0.24,  
$\Omega_\Lambda = 0.76$,  $h$ = 0.72) increasing the number of nodes and the size of nodes (up to 96 cores and 624GB Ram). We present our scalability results in Figure \ref{fig:gad2}. The GCP infrastructure is based on standard ethernet connections, while for HPC applications the role of a low latency high throughput interconnect is crucial, as evident from Figure \ref{fig:gad2}. On the other side, the cluster is suitable to any HTC applications where the inter-node communication is not present or limited.

\section{POC 3 - Workflow execution. Running GIANO-B data reduction pipeline as a service}
 In this use case we report and comment about the creation of a scaled and balanced environment, whose purpose is the execution of workflows submitted by the user through the workflow environment \text{\textit{Yabi}} \citep{yabi}. This scenario involves the user, who retrieves GIANO-B raw data from TNG archive public and private storage, and the execution of the GOFIO data reduction pipeline \citep{rainer} to produce reduced data that can be retrived by the user himself. The main aims of this PoC involves the simplification of the management of the infrastructure, moving from an on-premises infrastructure to PaaS/SaaS layers offered by GCP and of the deployment of software using containers to avoid incompatibility issues between packages that must coexist and work together. Finally, this PoC aims to improve software and service maintenance while optimizing and balancing the scalability of the service according to the load.
The implementation of this PoC foresees these services from GCP: Google Compute Engine for virtual machine instances management, Slurm or Google Kubernetes Engine as workload manager to deploy GOFIO container and the Docker platform for the containerization of GOFIO pipeline. Since we have two workload managers, two different solutions for this PoC was implemented. In the first architecture we made use of \textit{Yabi} and Slurm while in the second one we exploit \textit{Yabi} coupled with Kubernetes.
For both architecture \textit{Yabi} was deployed on a Compute Engine instance that acts as frontend for final user.
Slurm cluster was deployed using standard \textit{yaml} file available through Slumr official documentation\footnote{https://github.com/SchedMD/slurm-gcp}. To connect \textit{Yabi} with Slurm, we used the native \textit{Yabi}-Slurm backend connector, which is available in the latest version of \textit{Yabi} (version 9). Kuberntes cluster was deployed using Kubernetes Engine following the official Google documentation\footnote{https://cloud.google.com/kubernetes-engine/docs/how-to/creating-a-cluster}
deploying a Network File System (NFS) server from Cloud Launcher, configuring Persistent Volumes, POD ReplicaSet, LoadBalancer and HorizontalAutoScaler. \textit{Yabi} does not provide a default backend connector for Kubernetes, therefore we used the default \textit{Yabi}-SSH backend connector to connect \textit{Yabi} to Kubernetes cluster generating SSH key in \textit{Yabi} instance and adding it in the Kubernetes cluster. To test the performance of both architecture and to check actual scalability as function of the load, massive tests submitting simultaneously tens of jobs were performed. As a reference, for on-premises infrastructure these large workloads result in an excessive dilation of the execution times, since the total execution time of all the jobs (submitted simultaneously) is much greater than the sum of the execution times of the individual jobs performed one by one and, in most extreme cases, \textit{Yabi} crashes. For the architecture \textit{Yabi}-Slurm deployed on GCP the scalability is good and all jobs are completed correctly with no significant time leaks compared with the execution time of a single job. Slurm is natively supported by \textit{\textit{Yabi}} and it performs reasonably good in managing the job queue and the scaling. New Compute Engine instances are created and destroyed on demand efficiently according to the load.
However, for what concerns the \textit{\textit{Yabi}}-Kubernetes the scalability is also remarkable, but some jobs (about 1 each 8) exit with error and they are not more recovered, probably due to the fact that in this configuration the job queue is completely managed by the \textit{\textit{Yabi}} SSH Backend that submit jobs to Kubernetes which seems able to manage the load, but the \textit{\textit{Yabi}} Backend fails to manage all the job queue. We evaluate a total estimated charges of about 200 EUR/month to maintain both architectures up and running.

\bibliography{P6-8}
\end{document}